# A Dynamic Motion Planning Framework for Autonomous Driving in Urban Environments

Yuncheng Jiang[1,2], Xiaofeng Jin[1], Yanfei Xiong[1], Zhaoyong Liu[1]

1. Global Technology Co, Ltd. Shanghai 201800, China
E-mail: jiangyuncheng@glb-auto.com

2. Hongkong University of Science and Technology, Hongkong
E-mail: yjiangbw@connect.ust.hk

**Abstract:** we present a framework for robust autonomous driving motion planning system in urban environments which includes trajectory refinement, trajectory interpolation, avoidance of static and dynamic obstacles, and trajectory tracking. Given road centerline, our approach smoother the original line via cubic spline. Fifth order Bezier curve is then used to generate more human-like trajectories that guarantee at least second order continuity and curvature continuity. Dynamic trajectory planning task is decoupled into lateral spatial and longitudinal velocity planning problems. A bunch of candidate trajectory sets are generated and evaluated by an object function which considers kinematic feasibility, trajectory smoothness, driving comfort and collision-checking. Meanwhile, an LQG controller is used to generate longitudinal velocity profile to ensure safety and comfort. After that, spatial and velocity profiles are transformed into commands executed by lateral steering and longitudinal acceleration controllers. This framework is validated within a simulation study and has been deployed on our autonomous vehicle shown in Fig.1 that has travelled thousands of miles in urban environments.

## 1 Introduction

Autonomous vehicle has become a hot research topic in the past few decades. Basically, an autonomous vehicle consists of four subsystems: perception system, localization system, decision system, and control system [1]. Based on sensor fusion methods, perception system detects dynamic traffic via Lidar, Radar, camera, etc. Localization system provides accurate vehicle position by GPS navigation or Simultaneous Localization and Mapping (SLAM). The main functionality of decision system is to generate global routes, make vehicle behavior decisions and generate feasible trajectories dynamically according to the information from perception system. The control system calculates desired steering wheel angle and vehicle longitudinal acceleration so that the vehicle accurately tracks the desired trajectory.

The decision system design can be further divided into three subsystems: mission planning, path planning and path tracking. They are hierarchically layered to achieve the goal of motion planning. In this paper, the main focus is on pathing planning and path tracking.

In path planning system, the original reference waypoint is given by mission planning system, but the waypoint cannot be directly used due to its discontinuity. In our approach, cubic spline functions are used to generate new continuous polynomials, based on the original road centerline. We also present a fifth order Bezier curve control points generation method that generate control points based on given waypoints. The control points are then used to generate a smooth Bezier curve that has good continuity property [2].

Dynamic trajectory planning is divided into lateral spatial and longitudinal velocity planning. In lateral spatial planning, optimal control method is applied to achieve an optimal lateral trajectory in Frenet coordinate [3], while in longitudinal direction, to facilitate computation efficiency, an Adaptive Cruise Control (ACC) controller is used for velocity control and distance control. Applying optimal control approaches to pathing planning system is not new. The general process of lateral spatial planning based on optimization method is to generate feasible cost functional, and then minimize it under some constraints. While the criteria of choosing a cost functional is compliance with the principle of Bellman's optimality, the trajectory generated with minimum cost function should be similar to driving behavior of what human being will do. Therefore, in lateral spatial planning, this paper first elaborates a well-defined cost function which takes into consideration of trajectory continuity, comfort, safety (collision-free), kinematic feasibility and trajectory compliance between two iterations, etc. Then, to ensure instantaneity, the cost functional minimization is solved by KD tree, lazy collision-checking and exhaustive search [4][5]. In longitudinal velocity planning, ACC controller using LQR is used to generate comfortable and safe velocity profile. Kalman filter is used to deal with environment noise.

Another very important issue is path tracking. Basically, there are two difference methods in vehicle control: optimal control methods that rely on vehicle dynamics and geometric methods that rely on vehicle kinematics. In this paper, we use the Stanley model [6], and make some modifications to improve its system stability and performance.

The main contribution of this work is a comprehensive trajectory smoother, motion planning and trajectory tracking framework for autonomous driving that includes the following features:

- Cubic spline functions are used to generate new continuous polynomials, based on the original road centerline, consisting of uniformly distributed waypoints.
- A fifth order Bezier curve-based trajectory smoother is put forward which guarantees curvature continuity and driving comfort.

- Path planning is decoupled into lateral and longitudinal design, optimal control method is used to solve a functional minimization problem in lateral direction, and an ACC controller is used to achieve velocity and distance control in longitudinal direction.
- A kinematic model-based trajectory tracking controller is used that allows for the vehicle to track the desired reference line precisely.

The rest of this paper is organized as follows: Section II proposes a spline interpolation method and Bezier curve smoother. In Section III, pathing planning design is decoupled into lateral and longitudinal direction and solved separately. Kinematic model is introduced in Section IV to execute precise trajectory tracking. Section V is the conclusion.

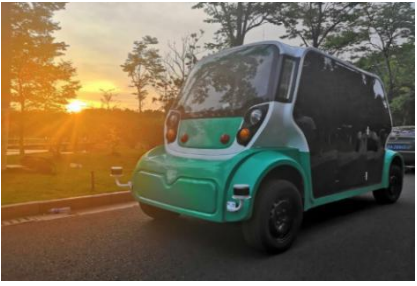

Fig. 1: The Autonomous Vehicle of Global Technology Co, Ltd. in Shanghai, China

## 2 Reference Line Generation

In this section, path model is proposed to generate original road centerline. The centerline will be interpolated so that the waypoints on the centerline is uniformly sampled. Then due to its discontinuity, a Bezier curve smoother is used to regenerate humanlike reference path.

### 2.1 Path Model

The path model used in autonomous driving is a pre-stored digital map. Road information is presented in this two-dimension map including global coordinates, road width, speed limit, traffic lights position, etc. Basically, the original road centerline is a sequence of waypoints generated by connecting the center waypoints of road [7]. However, the generated centerline is usually not uniformly distributed and may even have high-curvature bumps. Therefore, to generate a feasible reference line, we have to go through the two steps as follows:

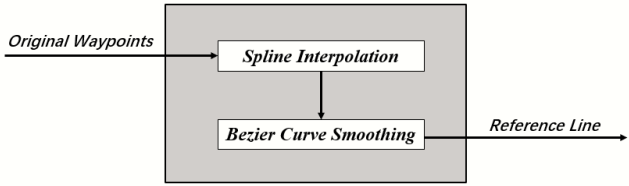

Fig. 2: Path Model. The original centerline is modified by interpolation and smoothing

### 2.2 Spline Interpolation and Bezier Curve Smoother

To interpolate the original waypoints into a uniformly distributed centerline, we introduce a monotonically increasing variable "***segment***" ($s_i$) to indicate the segments along the interpolated ceterline. Cubic spline functions are used to generate new waypoints with uniform distance between every two adjcent waypoints. Let's define the original centerline points as $(x_0, y_0), (x_1, y_1), \cdots, (x_n, y_n)$, and each segment $s_i$ is confined by its starting point $(x_i, y_i)$ and end point $(x_{i+1}, y_{i+1})$. The cubic spline defined on each segment $s_i$ is as follows:

$$f_i(x) = a_i + b_i(x - x_i) + c_i(x - x_i)^2 + d_i(x - x_i)^3 \quad (1)$$

The close form solution for each cubic spline parameters are:

$$a_i = y_i \quad (2)$$

$$b_i = \frac{a_{i+1} - a_i}{a_i} - \frac{h_i}{3}(c_{i+1} + 2c_i) \quad (3)$$

$$c_i = \frac{m_i}{2} \quad (4)$$

$$d_i = \frac{c_{i+1} - c_i}{3h_i} \quad (5)$$

where $h_i = x_{i+1} - x_i$ and $m_i$ can be solved numerically shown in [8].

The output of the cubic spline function is smoothed and uniformly distributed road centerline. One example is shown in Fig.3. Independent cubic splines are defined in each segment.

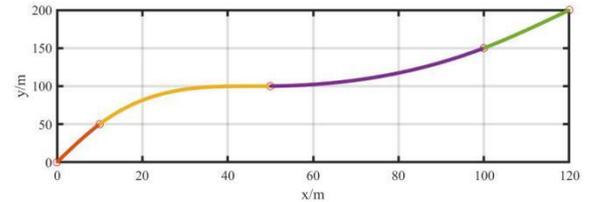

Fig. 3: Cubic Spline Interpolation

The interpolated centerline, however, may has abrupt curvature bumps at crossroads, which creates un-humanlike steering jerks when the autonomous vehicle is tracking the road centerline. We then introduce a Bezier curve smoother to guarantee that the new reference line after smoothing has at least second order continuity, and continuous curvature.

The Bezier curve of degree $n$ is represented as follows:

$$B(t) = \sum_{i=0}^{n} \binom{n}{i}(1-t)^{n-i} t * P_j) \quad (6)$$

where $t$ is a variable $\in (0,1)$, $j$ is the number of control points starting from $0$, and $B$ is the generated control points. In real cases, we will modify $t$ as follow:

$$t = \frac{p_c - p_s}{p_e - p_s} \quad (7)$$

where $p_c$ is current position, $p_s$ is starting position, and $p_e$ is end position. Fig.4 shows an example of how Bezier curve smoother works in a sharp intersection. The control points are generated as follows:

$$P_i = w_1^j + (w_2^j - w_1^j) * \frac{m_1 - d_i}{m_1} \quad (i=0,1,2) \quad (8)$$

$$P_i = w_2^j + (w_3^j - w_2^j) + \frac{d_{6-i}}{m_2} \quad (i=3,4,5) \quad (9)$$

where $d_1$, $d_2$, $d_3$ are fixed in $3,3$ and $8$ meters in this urban intersection scenarios [9]. Note that the value of $d$ is determined by current vehicle velocity, vehicle width and length, the shape of the intersection, and the road width. $m_1$ and $m_2$ are defined as:

$$m_1 = \|w_1, w_2\|_2 \quad (10)$$
$$m_2 = \|w_2, w_3\|_2 \quad (11)$$

Based on the control points $P$, a new Bezier curve **reference line** is generated, which is the output of path model.

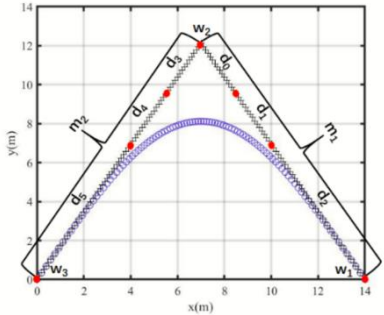

Fig. 4: Path Generation with Bezier Curve. In this sharp crossroad example, the interpolated road centerline is presented by black stars with uniform intersection of 0.01m. The control points of Bezier curve ($P_0$, $P_1$, $P_2$, $P_3$, $P_4$, $P_5$), presented by red dots, are calculated in Cartesian coordinates. The generated reference line is presented by blue circles.

Fig. 5, Fig. 6, Fig .7, respectively, present first-order derivative, second-order derivative and curvature continuity of the reference line.

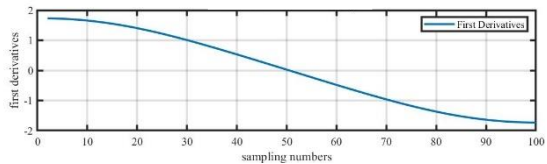

Fig. 5: First Order Derivative Continuity.

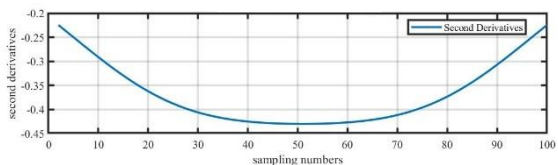

Fig. 6: Second Order Derivative Continuity.

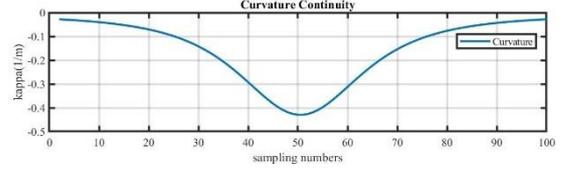

Fig. 7: Curvature Continuity.

## 3 Dynamic Trajectory Optimization

In this section, rather than formulating the trajectory optimization problem in Cartesian Coordinates, we use the Frenet Frame for trajectory generation and optimization [10]. Trajectory optimization is decoupled into spatial and velocity optimizers. In spatial optimization, cost function is first described and then solved by exhaustive search. To save computation expense, KD tree and lazy collision-checking are applied. Velocity optimization is solved by an ACC controller where an LQR controller is used, and environment noise is filtered via a Kalman filter.

### 3.1 Cost function

We define some cost functions to evaluate the safety (vehicle kinematics), human comfort, driving efficiency, energy consumption, trajectory consistency for each candidate trajectory [11]. We extract $n$ points from each trajectory and represent cost function in discrete forms, which are shown in Table 1.

Table 1: Cost Functions

| Cost | Formula | Physical Interpretation |
|---|---|---|
| $w_s$ | $\sum_{i=0}^{n} \|s_i\|^2$ | $s_i$ is path length of each section |
| $w_k$ | $\sum_{i=0}^{n} \|k_i\|^2$ | $k_i$ is curvature |
| $w_k$ | $\sum_{i=0}^{n} \|\dot{k}_i\|^2$ | $\dot{k}_i$ is first-order derivative of curvature |
| $w_k$ | $\sum_{i=0}^{n} \|\ddot{k}_i\|^2$ | $\ddot{k}_i$ is second-order derivative of curvature |
| $w_k$ | $\sum_{i=0}^{n} \|\dddot{k}_i\|^2$ | $\dddot{k}_i$ is third-order derivative of curvature |
| $w_{d_{center}}$ | $\sum_{i=0}^{n} \|d_{center}(i)\|^2$ | $d_{center}$ is lateral offset with the closest reference line |
| $w_{a_{lat}}$ | $\sum_{i=0}^{n} \|a_{lat}(i)\|^2$ | $a_{lat}$ is lateral acceleration |
| $w_{a_{lon}}$ | $\sum_{i=0}^{n} \|a_{lon}(i)\|^2$ | $a_{lon}$ is longitudinal acceleration |
| $w_{\dot{a}_{lat}}$ | $\sum_{i=0}^{n} \|\dot{a}_{lat}(i)\|^2$ | $\dot{a}_{lat}$ is rate of change of $a_{lat}$ |
| $w_{\dot{a}_{lon}}$ | $\sum_{i=0}^{n} \|\dot{a}_{lon}(i)\|^2$ | $\dot{a}_{lon}$ is rate of change of $a_{lon}$ |
| $w_t$ | $\sum_{i}^{n} \|l(s_i) - l_p(s_i^*)\|^2$ | $l(s_i)$ is distance between current trajectory and the reference line, $l_p(s_i^*)$ is distance between previous trajectory and the reference line |
| $w_t$ | $\sum_{i=0}^{n} \|t_i\|^2$ | $t_i$ is time duration of a trajectory |

The total cost for one trajectory is the weighted sum of all terms:

$$C_{total} = w_s \sum_{i=0}^{n} |s_i|^2 + w_k \sum_{i=0}^{n} |k_i|^2 + \cdots\cdots$$
$$+ w_l \sum_i^n \left| l(s_i) - l_p(s_i^*) \right|^2 + w_t \sum_{i=0}^{n} |t_i|^2 \quad (12)$$

In consideration of human comfort, we tend to choose trajectories with longer length, because shorter trajectories can easily be affected by unpredictable environment noise, and it is less sensitive to obstacle ahead. On the other hand, in consideration of efficiency, we also introduce $t_i$ to penalize longer trajectories that take more time to go through. In terms of trajectory smoothness and vehicle kinematic and dynamic constraints, we introduce curvature $k_i$, lateral acceleration $a_{lat}$, and longitudinal acceleration $a_{lon}$. To enhance the trajectory continuity, curvature derivatives from the first to the third order $\dot{k}_i$, $\ddot{k}_i$, $\dddot{k}_i$ are also considered. In terms of comfort, we consider lateral and longitudinal jerks $\dot{a}_{lat}$ and $\dot{a}_{lon}$. generally, we always expect that the vehicle drives along the reference line, therefore, we introduce $d_{center}$ to penalize trajectories deviating from the reference line. Lastly, the selection of current trajectory should also consider the past path, therefore, $l(s_i) - l_p(s_i^*)$, the distance between points on the current trajectory and on the selected optimal trajectory in the last circle is used to denote the consistency between consecutive regenerations. Too much difference between consecutive optimized trajectories is also penalized.

## 3.2 Lateral optimization

Lateral optimizer takes charge of generating quintic polynomials as candidate trajectories and selecting a collision-free trajectory with lowest cost function [3]. By using sampling-based method, we choose the starting state of our optimization $C_0 = [d_0, \dot{d}_0, \ddot{d}_0, l_0]$ according to the previously calculated trajectory. For the optimization itself, we sample in a constrained range whose state is defined as $C_e = [d_e, d_e, \dot{d}_e, l_e]$. We do not expect lateral acceleration and jerk, therefore, the end state is simplified as $C_e = [d_e, 0, 0, l_e]$. Practically, lateral offset $d_0$ and $d_e$ are confined in range $|d_0| \leq W_{road}$ and $|d_e| \leq W_{road}$ where $W_{road}$ is the width of road. For computation instantaneity, we also constrain the longitudinal search $l_e$ in range $L_{min} \leq l_e \leq L_{max}$ where $L_{min}$ and $L_{max}$ are minimum and maximum sampling lengths in longitudinal direction. Since cost functions are calculated in discrete form, sampling density of $d_e$ and $l_e$ are specified as $\Delta d_e = 0.1m$ and $\Delta l_e = 0.2m$ pragmatically. When given starting point configuration $C_0 = [d_0, \dot{d}_0, \ddot{d}_0, l_0]$, a new candidate trajectory is generated by sampling different end point configurations: $C_e = [d_e, 0, 0, l_e]$. The lateral optimization problem, therefore, can be mathematically expressed as follows:

$$\boldsymbol{min} \ \mathrm{C}_{total} = w_s \sum_{i=0}^{n} |s_i|^2 + \cdots + w_t \sum_{i=0}^{n} |t_i|^2$$

$$\boldsymbol{s.t.}$$
$$traj := traj\big(C_0 = [d_0, d_0, \dot{d}_0, l_0], C_e = [d_e, d_e, \dot{d}_e, l_e]\big)$$

$$L_{min} \leq l_e \leq L_{max}$$
$$|d_0| \leq W_{road}$$
$$|d_e| \leq W_{road}$$

Since the quintic polynomial has six constraints: $C_0 = [d_0, \dot{d}_0, \ddot{d}_0, l_0]$ and $C_e = [d_e, 0, 0, l_e]$, we can get a close form solution for the polynomial parameters. At the starting point $P(l_0)$, we have:

$$P(l_0) = \alpha_0 + \alpha_1 l_0 + \alpha_2 l_0^2 + \alpha_3 l_0^3 + \alpha_4 l_0^4 + \alpha_5 l_0^5 \quad (13)$$
$$P(\dot{l}_0) = \alpha_1 + 2\alpha_2 l_0 + 3\alpha_3 l_0^2 + 4\alpha_4 l_0^3 + 5\alpha_5 l_0^4 \quad (14)$$
$$P(\ddot{l}_0) = 2\alpha_2 + 6\alpha_3 l_0 + 12\alpha_4 l_0^2 + 20\alpha_5 l_0^3 \quad (15)$$

At the end point $P(l_e)$, we have:

$$P(l_e) = \alpha_0 + \alpha_1 l_e + \alpha_2 l_e^2 + \alpha_3 l_e^3 + \alpha_4 l_e^4 + \alpha_5 l_e^5 \quad (16)$$
$$P(\dot{l}_e) = \alpha_1 + 2\alpha_2 l_e + 3\alpha_3 l_e^2 + 4\alpha_4 l_e^3 + 5\alpha_5 l_e^4 \quad (17)$$
$$P(\ddot{l}_e) = 2\alpha_2 + 6\alpha_3 l_e + 12\alpha_4 l_e^2 + 20\alpha_5 l_e^3 \quad (18)$$

For convenience, let's assume that $l_0 = 0$. The six parameters can be solved as follow:

$$\alpha_0 = P(l_0) \quad (19)$$
$$\alpha_1 = P(\dot{l}_0) \quad (20)$$
$$\alpha_2 = \frac{P(\ddot{l}_0)}{2} \quad (21)$$

$\alpha_3$, $\alpha_4$, and $\alpha_5$ can be achieved by solving the following matrix function:

$$\begin{bmatrix} L^3 & L^4 & L^5 \\ 3L^2 & 4L^3 & 5L^4 \\ 6L & 12L^2 & 20L^3 \end{bmatrix} \times \begin{bmatrix} \alpha_3 \\ \alpha_4 \\ \alpha_5 \end{bmatrix} =$$
$$\begin{bmatrix} P(l_e) - (P(l_0) + P(\dot{l}_0)L + \frac{1}{2}P(\ddot{l}_0)L^2 \\ P(\dot{l}_e) - (P(\dot{l}_0) + P(\ddot{l}_0)L) \\ P(\ddot{l}_e) - P(\ddot{l}_0) \end{bmatrix} \quad (22)$$

where
$$L = l_e - l_0$$

Another important issue is collision-checking. Several circles are used to cover the area of the vehicle. If the distance from the circle to the obstacle is smaller than some safety threshold, the candidate trajectory is deleted. If we are lucky, the trajectory with lowest cost will pass collision-checking test, if it is not, the "second best" candidate trajectory will be tested, and so on. Even though we can use KD-tree to speed up distance finding, we still have to go through collision-checking process for all candidate trajectories if we introduce collision-checking term in cost function. Instead, to reduce computation expense, we propose lazy-collision-checking method, in which collision-checking is applied after the initial trajectory optimization, and some suboptimal collision-free trajectory can be found after testing only a small number of trajectories. One particular example is shown in Fig. 8. In this example, we illustrate how the autonomous vehicle, driving on its predefined reference line, avoid a moving obstacle in its way.

In each regeneration circle, optimal trajectory generation is shown in detail in Fig. 9.

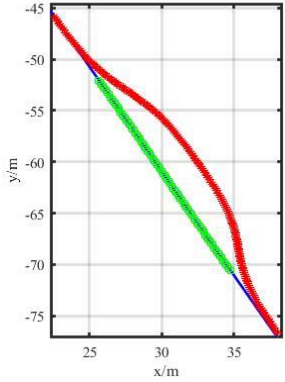

Fig. 8: Dynamic Obstacle Avoidance. The blue line is the reference line, the green square trajectory is front moving obstacles and the red circle trajectory is dynamically generated optimal trajectory to overtake the front obstacle.

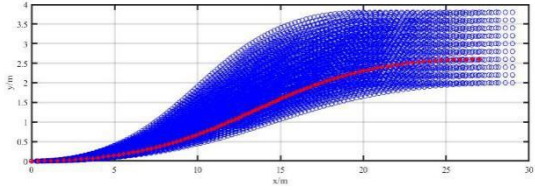

Fig. 9: Lateral Optimization Result. In this example, we specify $d_0 = 0$, $d_e \in [2,4]$, $\Delta d_e = 0.1m$, $\Delta l_e = 0.2m$, $15m \leq l_e \leq 30m$. The blue dot is generated candidate trajectories, and the red plus is the optimal trajectory that pass collision-checking test. In each regeneration circle, the optimal trajectory is selected as the input of trajectory tracking system.

### 3.3 Longitudinal optimization

An ACC controller takes charge of vehicle longitudinal control [12]. ACC controller can be divided into upper controller and lower controller. Upper controller calculates desired acceleration based on current traffic, and lower controller translates the desired acceleration into brake and throttle so that vehicle actuator can execute. In this section, we focus on the upper controller design. The state of the ACC controller $S_{ACC}$ is denoted as follows:

$$S_{ACC} = \{v_f, a_f, d, d_{desire}, v_p, a_p, a_{f,des}\} \quad (23)$$

the preceding vehicle is denoted by lowercase $p$, and the following vehicle is denoted by lowercase $f$. The inputs of upper controller are following vehicle velocity $v_f$, following vehicle acceleration $a_f$, real distance between two vehicles $d$, desired distance $d_{desire}$, preceding vehicle velocity $v_p$, and preceding vehicle acceleration $a_p$. The output of upper controller is desired acceleration $a_{f,des}$. The input of lower controller is desired acceleration $a_{f,des}$, and it is translated to brake and throttle as outputs. The objective of ACC system is to minimize the error between desired distance and real distance, relative velocity, and acceleration of the following vehicle.

for the sake of simplicity, desired distance with fixed headway time $\tau_h$ and constant safe distance $d_0$ are used:

$$d_{desire} = \tau_h v_f + d_0 \quad (24)$$

The lower controller and vehicle are combined together and simplified as a first order system with time constant $T_L$ and gain $K_L$. Time constant $T_L$ and gain $K_L$ well simulate real vehicle dynamics which always has a time delay and difference between real acceleration and desired acceleration. The relationship between desire acceleration $a_{f,des}$ calculated by upper controller and real acceleration $a_f$ actuated by lower controller and vehicle is as follows:

$$a_f = \frac{K_L}{T_L s + 1} a_{f,des} \quad (25)$$

To rewrite the vehicle dynamic system in a state space form, distance error $d_{error} = d_{desire} - d$, relative velocity $v_{rel} = v_p - v_f$ and acceleration of following vehicle $a_f = \frac{d}{dt} v_f$ are chosen as the three states which is written as $\dot{x} = Ax + Bu + \Gamma w$ where

$$x = \begin{bmatrix} d_{error} \\ v_{rel} \\ a_f \end{bmatrix}, A = \begin{bmatrix} 0 & -1 & \tau_h \\ 0 & 0 & -1 \\ 0 & 0 & -\frac{1}{T_L} \end{bmatrix}, B = \begin{bmatrix} 0 \\ 0 \\ \frac{K_L}{T_L} \end{bmatrix}, \Gamma = \begin{bmatrix} 0 \\ 1 \\ 0 \end{bmatrix},$$
$$w = a_p$$

Preceding vehicle acceleration $a_p$ is seen as disturbance to the system, and the output $y$ is the three states defined above:

$$y = Cx \quad (26)$$

where

$$C = \begin{bmatrix} 1 & 0 & 0 \\ 0 & 1 & 0 \\ 0 & 0 & 1 \end{bmatrix}$$

In real cases, the controller is designed in discrete form. The continuous state space is therefore discretized, by zero order hold method, into discrete form with sampling time $T$:

$$x_{k+1} = f(x_k, u_k, w_k) \quad (27)$$

which is supposed to have equilibrium state at $x = 0$, $u = 0$, and $w = 0$. In order to simplify this nonlinear problem, the system is linearized at equilibrium state, and rewritten as:

$$\tilde{x}_{k+1} = A\tilde{x}_k + Bu_k + g_k \quad (28)$$

By checking the observability matrix of $A$ and $C$, system states $\tilde{x}$ are ensured observable and measurable (full rank). System inputs are constrained in a safe range. According to national regulation on ACC system, the input $u$ is constrained in a range of $[-0.25g, 0.25g]$. Process noise $g$ is considered as Gaussian white noise with zero mean.

After linearization, the problem of search for an optimal feedback controller can be seen as a linear quadratic optimization problem. The control objective, as mentioned in previous section, is to minimize the error between desired distance and real distance, relative velocity, and acceleration of the following vehicle. Even though the cost function, theoretically, is infinite horizon, a finite yet long enough time period is chosen for feasibility of cost function calculation in real cases. The cost function, therefore, in discrete form, is given:

$$J = \frac{1}{N} \sum_{k=0}^{N-1} [x^T(k)Qx(k) + u^T(k)Ru(k)] \quad (29)$$

where Q matrix is positive semi-definite, and R matrix is positive definite. The two weighting matrices are written as:

$$Q = \begin{bmatrix} \rho_1 & 0 & 0 \\ 0 & \rho_2 & 0 \\ 0 & 0 & \rho_3 \end{bmatrix}, R = [r]$$

The optimal output is then written in the form of $u = -Kx$, where $K = R^{-1}B^T P$. $P$ can be obtained by solving Algebraic Riccati Equation ($ARE$), which is written as:

$$A^T P + PA + PBR^{-1}B^T P + Q = 0 \quad (30)$$

Once the Q and R matrices are determined, optimal $u$ can be achieved. Only when the states are perfectly measured, the optimal $u$ is globally optimal. In real scenarios, however, due to the process gaussian noise $g$, the optimal $u$ is only locally optimal. In order to compensate for the process noise effect, the input states of upper controller should be well estimated and filtered by Kalman Filter (**KF**).

As shown in Fig. 10, state $y$ is first filtered by Kalman filter and then be used. *LQR* with **KF** is what we call **LQG** controller.

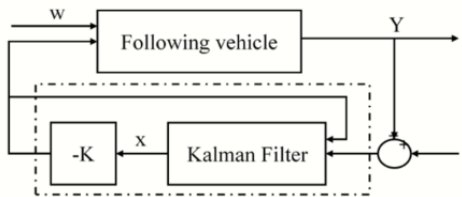

Fig. 10: Structure of LQG controller

## 4    Trajectory Tracking

The goal of trajectory tracking is to (1) minimize the lateral offset between the vehicle and the closest reference line, (2) minimize the different of heading angle between vehicle and reference line, and (3) smooth steering wheel control while maintaining stability.

In this section, we use the Stanley model as the lateral controller. Vehicle kinematics is first introduced. To improve the model stability, we then propose some modifications to Stanley model where the model parameters vary with vehicle velocity.

### 4.1    Vehicle Kinematics

The Stanley model, one of the famous geometric path tracking models, is first proposed by Stanford University's autonomous vehicle entry in the 2007 DARPA Grand Challenge, Stanley [13]. In this paper, a bicycle model based on Ackerman steered vehicle is used for simplification.

As shown in Fig. 11, the cross-track error between the center of the front axle to the nearest waypoint of the reference line is denoted by $e_{fa}$, and the difference heading angle between vehicle and nearest waypoint ($c_x, c_y$) is denoted by $\theta_e$.

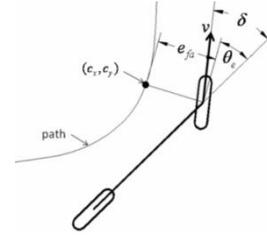

Fig. 11: Stanley Model Geometry

### 4.2    Lateral Control

In this section, we use Stanley model as the lateral controller. Vehicle kinematics is first introduced, and then we propose some modifications to Stanley model where the model parameters vary with vehicle velocity.

Stanley model is a feedforward and feedback function. The desired steering angle, denoted by $\delta$ consists of two parts:

$$\delta(t) = \theta_e(t) + tan^{-1}(\frac{ke_{fa}(t)}{v_x(t)}) \quad (31)$$

The first term is a feedforward control that compensate for angle difference between vehicle heading and trajectory heading. When the vehicle deviates from the desired trajectory, the second term adjust the steering angle such that the intended trajectory intersects the path tangent from ($c_x, c_y$). The function, however, does not guarantee its stability when the vehicle velocity is zero: the denominator of the second term goes to zero. We modify the control law as follow:

$$\delta(t) = \theta_e(t) + tan^{-1}(\frac{k_e(v_x(t))e_{fa}(t)}{L_x(v_x(t))}) \quad (32)$$

where $k_e$ and $L_x$ are functions of vehicle velocity.

It is feasible to set look forward distance as a function of vehicle velocity. With vehicle velocity increasing, a longer look forward distance avoid strong steering control when there is large $e_{fa}$. Meanwhile, we also find that Stanley has a singularity when vehicle velocity reaches zero, it is modified by replacing the denominator by another velocity related function $L_x$. Suppose that $L_x$ is defined as look forward distance, and it increases with velocity. In view of $k_e$, it also increases with velocity, as we find that vehicle lateral control is very sensitive to $k_e$ when velocity is low, while less sensitive as velocity increases.

In vehicle tests, we somehow find *(1)* relation between velocity and $k_e$, and *(2)* relation between velocity and $L_x$, that lead to good system performance. The two relations are defined as follow:

$$L_x = \begin{cases} 10, & v \in [0,12.5) \\ 0.8v, & v \in [12.5,25) \\ 20, & v \in [25,inf) \end{cases} \quad (33)$$

$$k_e = \begin{cases} 0.02v + 0.5, & v \in [12.5,25) \\ 1, & v \in [25,inf) \end{cases} \quad (34)$$

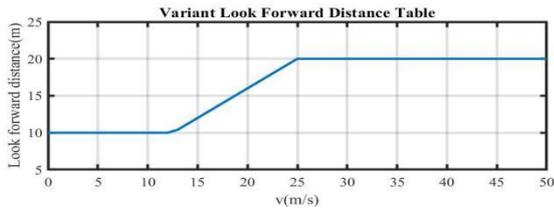

Fig. 12: Variant Look Forward Distance. It starts from a nonzero number 10 and reaches maximum at 20.

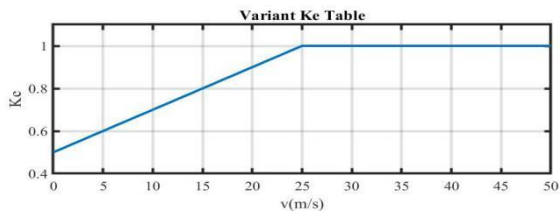

Fig. 13: Variable $K_e$. It starts from a constant number 0.5 and increases linearly until reaches its upper limit of 1.

In vehicle test, we test two typical scenarios to show system performance of trajectory tracking: (1) right turning at a crossroad, and (2) zig-zag curve tracking, which are respectively shown in Fig. 14.

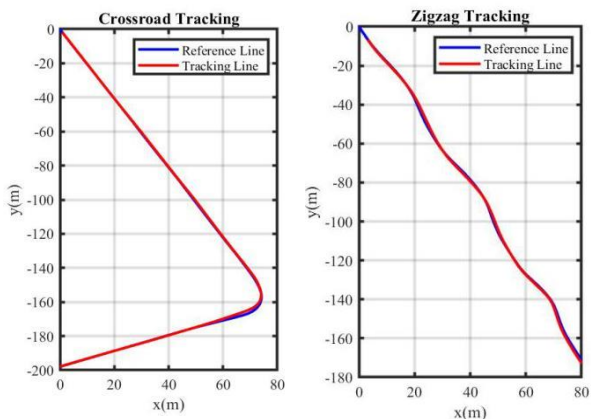

Fig. 14: Trajectory Tracking Results

## 5 Conclusion

This paper has proposed an efficient path planning framework for autonomous driving. In this framework, the original waypoints extracted from digit map is refined and interpolated via cubic spline refinement and fifth order Bezier curve interpolation, respectively. Path planner is divided into lateral spatial planning and longitudinal velocity planning. to save computation expense, KD tree, lazy collision-checking and exhaustive searching are used in optimal trajectory search. While in longitudinal direction, LQR with Karman Filter guarantees generating a safety and comfortable speed profile in real urban scenarios. Lastly, modification on the Stanley model ensures trajectory tracking stability and system performance.

For future research, high definition map should be used to provide more map semantic information. Obstacle tracking and prediction should also be further research to provide more accurate guide for path planning.